\documentstyle[10pt,epsfig,twocolumn]{ICEAA}

\title{\bf Optimization of the Coupling of High-Frequency Horn 
Antenna Array to the ESA PLANCK Submillimeter-Wave Telescope}
\author{Vladimir Yurchenko$^{(1, 2)}$,  John Anthony Murphy$^{(1)}$,  
Jean-Michel Lamarre$^{(3)}$}
\affiliation{$^{(1)}$Experimental Physics Department, 
National University of Ireland, Maynooth}
\address{Maynooth, Co. Kildare (Ireland)}
\affiliationb{$^{(2)}$Institute of Radiophysics 
and Electronics, National Academy of Sciences}
\addressb{12 Proskura St., Kharkov, 61085 (Ukraine)}
\affiliationc{$^{(3)}$Institut d'Astrophysique Spatiale, 
Universite de Paris XI}
\addressc{Bat. 121, Orsay Cedex, Paris, 91405 (France)}
\email{v.yurchenko@may.ie, \ amurphy@may.ie, \ lamarre@ias.u-psud.fr}

\begin{document}

\maketitle


\thispagestyle{plain}\pagestyle{plain}

\begin{abstract}
We study the electromagnetic coupling of the array of Gaussian and 
multi-moded horn antennas to the dual-reflector submillimeter-wave 
telescope on the ESA PLANCK Surveyor designed for measuring the 
temperature anisotropies and polarization characteristics of the 
cosmic microwave background. In this paper, we present the results 
of our analysis concerning the measurement of polarization with 
tilted off-axis dual-reflector Gregorian telescope and the 
propagation of multi-moded beams through such a system. 
\end{abstract} 

\section{Introduction} 

The dual-reflector submillimeter-wave telescope on the ESA PLANCK 
Surveyor is being designed for measuring the temperature 
anisotropies and polarization characteristics of the cosmic 
microwave background (CMB). 

PLANCK will carry two highly complex focal plane instruments, 
giving the wide frequency coverage ($30 GHz$ to $850 GHz$) 
and high sensitivity (of $\Delta T/T ~ 10^{-6}$) necessary 
to allow for the clean separation of the CMB from confusing 
foreground sources of radiation and the determination of the 
power spectrum of the CMB anisotropies on angular scales 
down to $5$ arcminutes. 

Our research is concerned with one of these key instruments, 
the Far-IR High Frequency Instrument (HFI), which will cover 
six frequency bands centered at $100$, $143$,$217$, $353$, 
$545$ and $857 GHz$ (wavelength range from $3 mm$ down to 
$350 \mu m$) [1]. At the telescope focal plane the HFI  
consists of an array of $36$ horn antenna structures (Fig.~1)
feeding the ultra sensitive bolometric detectors which will be 
cryogenically cooled to a temperature of $100 mK$.

One objective of our research is the optimization of the HFI 
optical design and the determination of the complete set of HFI 
beam patterns. The challenge of the problem is that the telescope 
is electrically large ($D/\lambda=4300$ at $\lambda=350 \mu m$) 
and consists of two essentially defocused ellipsoidal reflectors 
providing a very large field of view at the focal plane in order 
to accommodate the two focal plane instruments. 

The latter has a detrimental effect on the quality of the imaging 
introducing coma and astigmatism even at the center of the field 
of view where the highest frequency channels are placed. 
Furthermore the focal surface of least confusion has quite an 
unusual shape and it is extremely important to precisely place 
the phase centers of the HFI horn antennas on this surface.

Another objective is the characterization of the non-conventional 
polarization properties of the multi-beam telescope system. 
The CMB polarization is expected to be at a level of only 10\% of the 
temperature anisotropy quadrupole, and the success of the measurements 
will depend crucially on the precise knowledge of the polarization 
properties of the telescope.

\section{Formulation of the problem}

While the performance of the antenna structures can be thoroughly 
tested in terrestrial conditions, the coupling of the HFI with the 
telescope is, basically, optimized through the computer simulations. 
Such simulations are also needed for the data retrieval from the 
raw PLANCK measurements once in orbit. 

The most intricate part of the problem is the computation of the 
polarization patterns of the telescope beams from the linearly 
polarized horns and of the power patterns of multi-moded beams 
of extremely high frequencies (the beams of $12$ modes at 
$545 GHz$ and of $30$ modes at $857 GHz$).

Among various simulation techniques, physical optics (PO) is the 
most adequate one for the given purpose. However, conventional 
implementations of the technique [2, 3] do not fit the size of 
the problem. Commercially available packages are also very limited 
in their capacity to rigorously answer this sort of questions.
For example, even the best commercial software requires about 
one week of computation of the main beam of the telescope at the 
relatively low frequency of $143 GHz$, while all conventional 
physical optics codes collapse at the highest frequencies of 
$545 - 857 GHz$.

To solve the problem, we developed a special PO code [1] that 
allowed us to overcome the limitations of a generic approach for 
large multi-reflector systems and perform typical simulations of 
the telescope in the order of minutes. With recent improvements, 
it requires only $2$ minutes for the fairly accurate PO simulation 
of the telescope beam at the frequency of $143 GHz$ and about 
$30$ minutes for the beam of $30$ modes at $857 GHz$ using a 
PC Pentium III ($500 MHz$) under the Linux operating system. 

%
 \begin{figure}
\centerline{\includegraphics{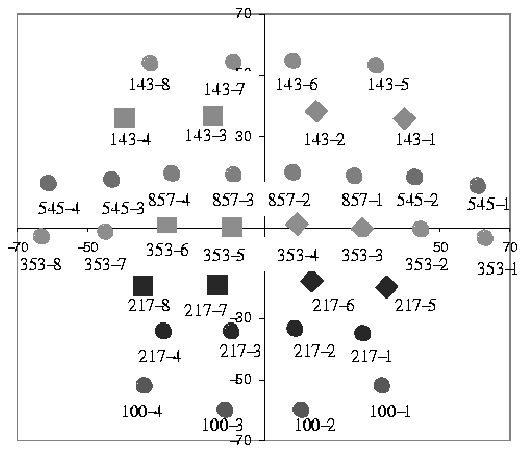}}

{\small Fig.1. Horn positions on the focal plane as seen from the 
secondary mirror ($X_{RDP}$ and $Y_{RDP}$ axes in the plane 
are directed upwards and to the left, respectively, while 
the scan direction over the sky is to the right)
}
 \label{fig1}
 \end{figure}
%

In this paper, we consider the beam of the Gaussian horn HFI-143-1 
designed for the separate measuring of the fields of two orthogonal 
linear polarizations, 'a' and 'b', which correspond to the electric 
field at the beam axis in the sky tilted with respect to the local 
vertical (defined below) by the angle $\psi_a =-45^{\circ}$ and 
$\psi_b =+45^{\circ}$, respectively (the angle $\psi$ is measured 
clockwise from the local vertical as seen from the telescope).

We also provide the results of our simulations of the defocusing 
effect of the $30$-mode horn HFI-857-1 from the old layout of 
the focal plane unit.

The electric field at the aperture of the Gaussian horn HFI-143-1 
is linearly polarized and specified by the formula
\begin{equation}
\vec E(r)=E_0 \exp(-r^2/w_a^2) \ \vec e, \quad  0\le r \le a
\end{equation}
where $r$ is the radial coordinate, $w_a =2.68mm$ is the beam 
waist at the aperture, $a=5.0mm$ is the aperture radius and 
$\vec e=\vec e_{a,b}$ is the unit polarization vector 
($\vec E(r)=0$ outside the aperture). 

Both the power and phase patterns of such a feed coincide 
perfectly well with the available experimental data. 
The far-field pattern of this particular feed satisfies the 
edge taper requirement of being below $-25 dB$ at the angles 
$\theta \ge 25^{\circ}$.

H-857-1 is a $30$-mode back-to-back conical corrugated horn~[4]  
with the electric field at the aperture specified by the set of 
hybrid modes similar to those used in [1] for the HFI-545-1 horn.
The total field is defined as a non-coherent superposition 
of all the modes of all polarizations.

Amplitudes of the aperture fields are normalized so that each 
polarized mode has the total power $P_m=1$.

%
 \begin{figure}
\centerline{\includegraphics{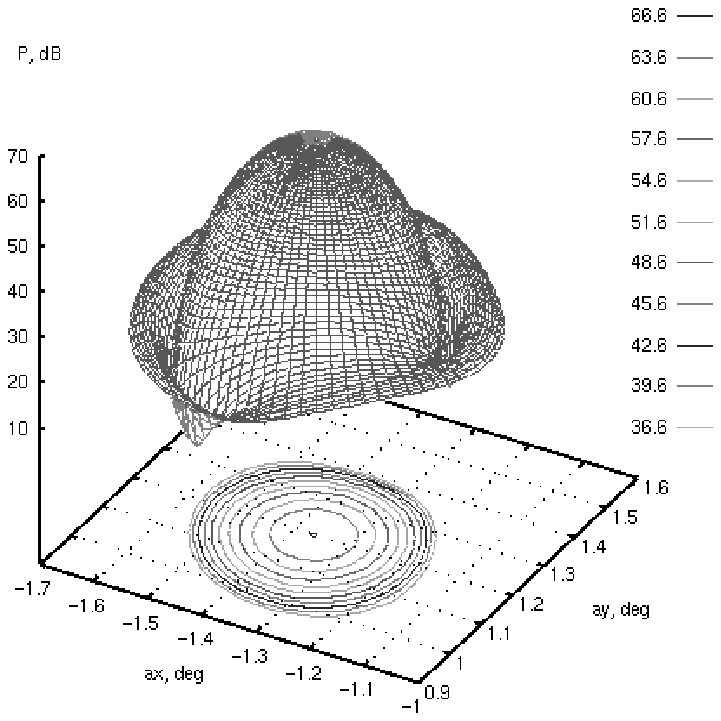}}

{\small Fig.2. Power pattern of the telescope beam HFI-143-1 
(power pattern is the same for both polarizations) 
}
 \label{fig2}
 \end{figure}
%

\section{Polarization of the Gaussian beams}

Fig.~2 shows the power pattern of the telescope beam H-143-1 as 
projected on the plane normal to the telescope line-of-sight 
at the $(0,0)$ point ($a_x$ and $a_y$ are the horizontal and 
vertical axes on the plane, respectively, measured in degrees). 
The beam axis is at the point $a_x= -1.365^{\circ}$, 
$a_y= 1.197^{\circ}$ which is defined as the point of maximum 
power of the beam. 

The beam is well shaped down to $-30dB$ below the maximum and 
can be approximated by a Gaussian function (at this level, 
the pattern does not depend on polarization) with a full beam 
width of $W_{min}=8.4$ arcmin and $W_{max}=9.2$ arcmin 
measured at $-3dB$.

The polarization of the beam is generally elliptical except 
precisely at the beam axis where it remains linear. In order 
to achieve the required orientation of the polarization pattern 
in the sky, we should orient the polarization vector $\vec e$ 
properly on the horn aperture~[1]. 

For immediate comparison of polarizations measured by different 
horns when scanning through the sky, we should use easily aligned 
directions in the sky as equivalent reference polarization axes 
for different beams. Such directions are the meridians in the 
spherical frame of the telescope, with the telescope spin axis 
being the pole (the meridians define local verticals at various 
observation points, while the parallels are local horizontals 
that constitute the orthogonal directions). 

%
 \begin{figure}
\centerline{\includegraphics{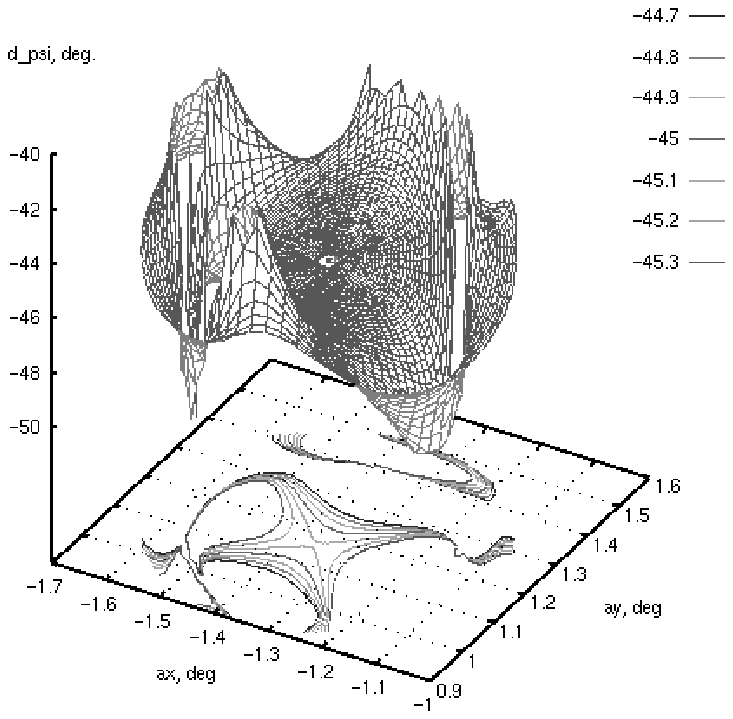}}
(a)
\centerline{\includegraphics{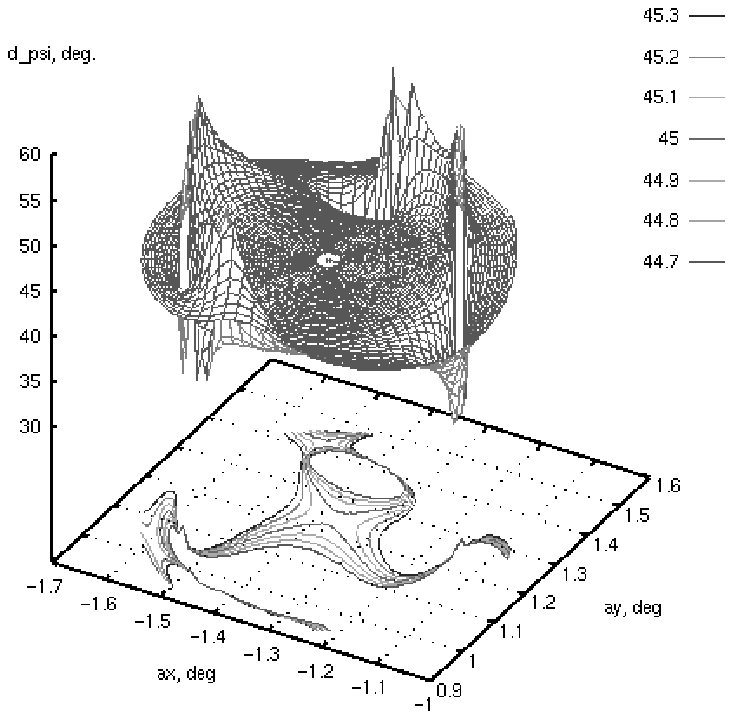}}
(b)

\vspace{2mm}

{\small Fig.3. Deviation of the major axis of polarization ellipse 
from local vertical for the telescope beams HFI-143-1a and 
HFI-143-1b with (a) $\psi_a =-45^{\circ}$ and 
(b) $\psi_b =+45^{\circ}$, respectively 
\vspace{-2mm}
}
 \label{fig3}
 \end{figure}
%
%
 \begin{figure}
\centerline{\includegraphics{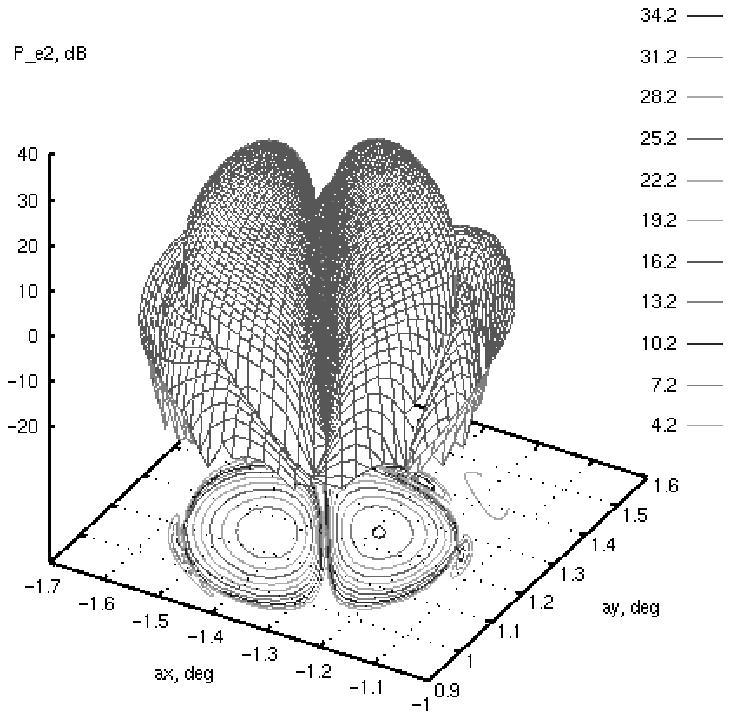}}

{\small Fig.4. Magnitude of the minor semi-axis of polarization ellipse 
in the far field of the telescope beam HFI-143-1a (both cases 
$\psi_a =-45^{\circ}$ and $\psi_b =+45^{\circ}$ are rather similar)
}
 \label{fig4}
 \end{figure}
%

Also, we should properly define the reference axis for the 
polarization vectors $\vec e$ of tilted horns. We define the 
reference axis as the direction of $\vec e$ in the horn aperture 
plane that is projected on the vertical axis of the focal plane. 
The orientation of $\vec e$ is specified by the angle $\phi$ 
in the horn aperture plane measured from this reference in a 
clockwise direction when looking from the horn to the secondary 
mirror. 

Using these definitions, we have found that the beam of the 
H-143-1 horn is polarized at its axis at the required angles 
of (a) $\psi_a = -45^{\circ}$ and (b) $\psi_b = +45^{\circ}$ 
if the horn polarization vector $\vec e$ is specified by the 
angle $\phi_a=-46.91^{\circ}$ and $\phi_b=+43.09^{\circ}$, 
respectively. 

In terms of the coordinate frame $M1$ of the primary mirror 
used as a global frame for the design specification, the unit 
polarization vectors found above are represented by their 
components as follows
\begin{eqnarray}
\vec e_a = (\ 0.522212,\ -0.727795,\ 0.444532 \ ) \nonumber \\
\vec e_b = (\ 0.628105,\ \ 0.680820,\ \ 0.376787 \ ).
\end{eqnarray}

The deviation of the major axis of the polarization ellipse from 
the local meridian as a function of the observation point 
within the beam in each of the two cases is shown in Fig.~3, (a)
and (b), respectively.

When the polarization vector $\vec e$ is properly oriented, 
the cross-polarized component of the far field measured along  
the respective orthogonal direction in the sky is minimized.
In such a case, the power pattern of the cross-polarized 
component is, basically, determined by the power pattern 
of the semi-minor axis of the polarization ellipse at each 
observation point of the beam, Fig.~4. 

The latter is very much the same for any orientation of the 
polarization vector and can be approximated as follows 
\begin{equation}
P_{cr}=P_{cr0} \ [(\theta/\theta_0) \ \sin(\varphi-\varphi_0)]^p \ 
\exp[(\theta/\theta_0)^q]
\end{equation}
where $P_{cr0}, \ dB$, is the maximum power of the minor axis 
component achieved at the points specified by the polar angle 
$\theta_0$ and the azimuthal angles $\varphi_0 \pm 90^{\circ}$ 
measured from the center of the pattern which is located at the 
beam axis (the values of the parameters in this approximation 
depend on the position of the horn considered).

\section{Simulation of multi-moded horns} 

Multi-moded horns are designed for receiving maximum microwave 
power within the required angular resolution consistent with the 
requirements on the beam taper at the primary mirror of a level 
of $-25dB$, with an aspect angle of $25$ degrees. These are 
rather restrictive and contradictory requirements resulting in the 
big aperture area of the corrugated conical horns specially 
optimized for the given application~[4].

Since the phase front of the aperture field of a conical horn 
is convex, the effective focal center is located inside the horn 
at a distance $R_C$ from the aperture. The horn position is 
specified by the aperture refocus parameter $R_A=R_F+R_C$ 
where $R_A$ is the distance from the focal plane to the horn 
aperture measured along the horn axis and $R_F$ is the similar 
distance to the geometrical focus on this axis as found by the 
ray-tracing software. 

One may try to make an estimate of $R_C$ by considering the 
basic mode of the horn as a Gaussian beam propagating through 
the horn aperture. For the horn HFI-857-1 (old layout) with 
the slant length $L=28mm$ and the aperture radius $a=2.55mm$, 
the estimate is $R_C=11.9mm$ while $R_F=9.1mm$ that results 
in $R_A=21.0mm$. 

Accurate simulation shows, however, that such an estimate is 
very misleading. When computing the gain $G$ defined as 
$G=10\log(4\pi P(0)/P_0)$ where $P(0)$ is the power of the 
total multi-moded field on the beam axis and $P_0$ is the 
radiated power of a single polarized mode, one can find that
$G$ as a function of the aperture refocus parameter $R_A$ has 
a maximum at $R_A=13.0mm$ (Fig.~5).

The accurate value of $R_A$ is different from the 
rough estimate above by $\Delta R_A =8mm =23\lambda$  
that results in an extra gain of about $2.5dB$.
Thus, the multi-moded horns should be placed slightly 
further from the secondary mirror compared to the 
positions found by simple estimates.

\section{Conclusions}

A fast physical optics simulator has been developed for the 
analysis of the dual-reflector submillimeter-wave telescope 
on the ESA PLANCK Surveyor. The code overcomes the limitations 
of a generic approach for large multi-reflector quasi-optical 
systems and can perform typical simulations of the telescope 
in the order of minutes. 

%
 \begin{figure}
\centerline{\includegraphics{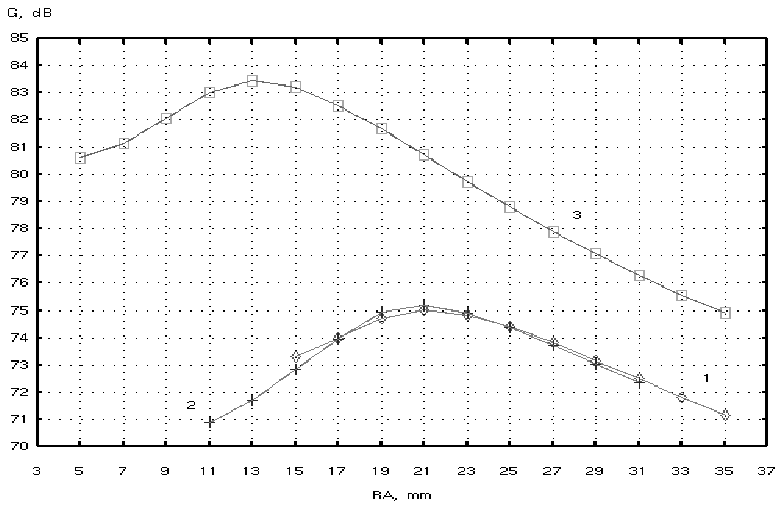}}

{\small Fig.5. Defocusing effect of the 30-mode conical corrugated horn 
HFI-857-1, old layout (1- Gaussian beam approximating the basic mode,
2- Basic mode, 3- Total field as a non-coherent superposition of 
$30$ modes)
}
 \label{fig6}
 \end{figure}
%

A study of the power patterns, polarization characteristics, 
defocusing effects and modal structure of the telescope beams 
from both Gaussian and multi-moded horns has been performed.
Analysis of the beams from the linearly polarized Gaussian horns 
have shown that the far-field of the telescope is, generally, 
elliptically polarized except precisely at the beam axis where 
linear polarization is preserved. 

When rotating the polarization vector of the horn field about 
the horn axis, the major axes of the polarization ellipses 
at the central part of the telescope beam rotate by very much 
the same angle about the direction of observation 
while the power patterns of both the co- and cross-polarized 
components of the far field remain basically unchanged. 
Systematic deviations from this basic rule which occur mainly 
in the peripheral part of the beam can be successfully simulated 
and taken into account.

Even for the most tilted Gaussian horns located at the edge 
of the horn array, the power associated with the minor axes 
of the polarization ellipses in the telescope beam remains 
at the level of more than $30dB$ below the maximum power 
of the beam. 

In order to achieve maximum angular resolution with multi-moded 
horns, the latter should be located slightly further from the 
secondary mirror compared to the estimates based on the Gaussian 
approximation of the basic mode applied to the optimal 
focal positions as found by the ray tracing techniques.

\section{Acknowledgement}

The authors are grateful to Yuying Longval for providing updated 
positions and aiming angles of the high-frequency horns along 
with the map of the focal plane. V.Y. and J.A.M. would like 
to acknowledge the support of Enterprise Ireland.


\begin{thebibliography}{99}

\bibitem{bib1} V. Yurchenko, J. A. Murphy, and J. M. Lamarre,
``Fast physical optics simulations of the multi-beam dual-reflector 
submillimeter-wave telescope on the ESA PLANCK surveyor'', 
{\em Int. J. Infrared and Millimeter Waves}, January 2001,
Vol.22, No.1, pp. 173-184.

\bibitem{bib2} L. Diaz and T. Milligan, ``Antenna Engineering 
Using Physical Optics: Practical CAD Techniques and Software'',  
London: Artech House, 1996.

\bibitem{bib3} C. Scott, ``Modern Methods of Reflector Antenna 
Analysis and Design'',
London: Artech House, 1990.

\bibitem{bib4} R. Colgan, J. A. Murphy, B. Maffei, C. O'Sullivan, 
R. Wylde, and P. Ade,
``Modelling Few-Moded Horns for Far-IR Space Applications'',
{\em The 11th Int. Symp. on Space Terahertz Technology, Ann Arbor,
Michigan}, May 2000, pp. 368-378.

\end{thebibliography}
\end{document}